\documentclass[11pt,twoside]{article}

\usepackage{graphicx}
\usepackage{asp2006}

\begin{document}
\title{The Effect of the ISM Model on Star Formation Properties in Galactic Discs}  
\author{Elizabeth J. Tasker\altaffilmark{1} and Greg L. Bryan\altaffilmark{2}}   
\altaffiltext{1}{Department of Astronomy, University of Florida, Gainesville, FL 32611}  
\altaffiltext{2}{Department of Astronomy, Columbia University, New York, NY 10027}

\begin{abstract} %%% Abstract to run on from here.
Modelling global disc galaxies is a difficult task which has previously resulted in the small scale physics of the interstellar medium being greatly simplified. In this talk, I compare simulations of galaxies with different ISM properties to determine the importance of the ISM structure in the star formation properties of the disc. 
\end{abstract}

\section{Introduction}

One of the main problems in understanding galaxy formation is that star formation is extremely complex. The gas out of which stars form is a turbulent mix of forces in which gravitational collapse, thermal pressure, magnetic fields, cosmic ray pressure and energy from supernovae all fight for dominance \citep{MacLow2004}. 

This leaves galaxy simulators with a choice; either to model the interstellar gas in detail but restrict their study to a small patch of the galaxy \citep[e.g.][]{Slyz2005} or to simulate the entire galaxy but use a greatly simplified model for the ISM \citep[e.g.][]{Li2005}. The former approach allows the inclusion of many more of the important physical processes but is unable to tell us anything about the global evolution of the disc. The latter, meanwhile, reveals properties of the whole galaxy including star formation histories and global structures, but it is impossible to judge the effect the simple ISM model is having on these results. 

However, recent numerical simulations are now able to include a more complex multiphase ISM in these global models \citep{Tasker2007, Tasker2006, Wada2007}. This allows us to test the impact of modelling the ISM in a more realistic way in galaxy formation. This has particular baring on cosmological simulations, where the resolution of the ISM of individual galaxies is still beyond our reach.  

\section{Numerical method}

Using the AMR code, {\it Enzo} \citep{Bryan1997}, we compared three models of isolated galaxy discs where we varied the properties of the interstellar gas. In all cases, the simulations started with a rotating Milky Way-sized exponential disc of gas sitting in a static NFW dark matter potential. The set-up is described in more detail in \citet{Tasker2006, Tasker2007}. In our first galaxy model (ISM~\#1), the gas was allowed to radiatively cool to 300\,K. In our second model (ISM~\#2), an additional background photoelectric heating term was included while in our last disc (ISM~\#3), the ISM was kept at a fixed temperature of 10,000\,K, in keeping with previous simulations. All simulations included star formation and ISM models 1 and 2 also included feedback from Type II SNe (this was impossible to include for the third, isothermal, disc).

\section{The structure of the ISM}

\begin{figure}
\begin{center}
\includegraphics[width=\textwidth]{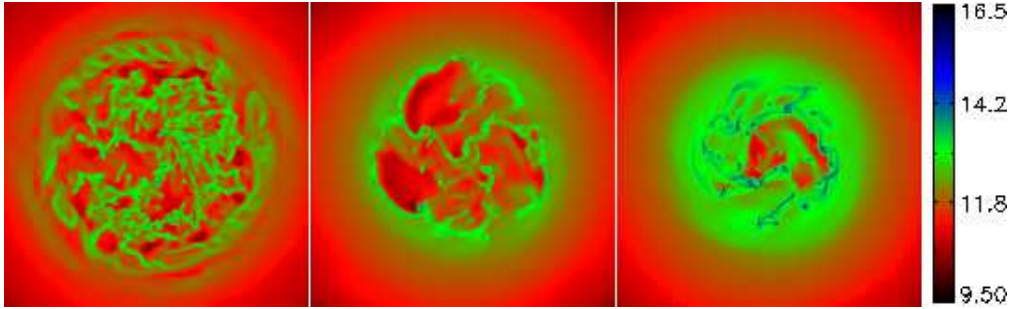}
\caption{Gas density projections at 388\,Myrs for (left to right) ISM~\#1, ISM~\#2 and ISM~\#3, all without feedback. Scale is to the base-10 logarithm with units M$_\odot$Mpc$^{-2}$ and each image is 60\,kpc across.}\label{fig:gasproj}
\end{center}
\end{figure}

Figure~\ref{fig:gasproj} shows projections of the gas density of each galaxy disc 377\,Myrs after the start of the simulation. At this time, the disc has undergone its initial fragmentation (and subsequent star burst) and the evolution is now slow. In Figure~\ref{fig:gasproj}, all images are for runs without stellar feedback. With our first ISM model, shown on the far left, we see that the gas has fragmented out to a well-defined radius. Beyond this point, there is still a significant amount of gas, but it is stable to gravitational fragmentation and does not collapse to form stars. When we include photoelectric heating in ISM~\#2 (middle image) we see a notably different gas structure. In particular, there exists large voids of hot, low density gas. The porous nature of this ISM has been observed both in our own galaxy and (perhaps most dramatically) in the HI map of the LMC. This result agrees with recent simulations of \citet{Wada2000}, who suggest these holes are not the results of SNe remnants, but rather the product of gravitational and thermal instabilities. Our final ISM model, where the temperature is fixed, shows another distinct structure. Due to the fixed high temperature of $10^4$\,K, the Jean's Length is higher than in the other discs, causing the star-forming knots of gas to be much higher in mass. This has the effect of producing very large star clusters that are confined to the densest, inner regions of the disc. Their formation also produces voids in the gas distribution, but this is due to gas deficit, not to instabilities.

The pressure distribution of these discs is also interesting. Without feedback, discs with ISM~\#1 and ISM~\#2 are largely in pressure equilibrium, as predicted by the analytical models of \citet{McKee1977}. The isothermal disc cannot be, since fixing the temperature requires the pressure to be proportional to the gas. When feedback is introduced, this situation changes. SNe energy causing streams of gas to be blown both in the disc's plane and off its surface, creating a galactic fountain. 

\begin{figure}
\includegraphics[angle=270, width=\columnwidth]{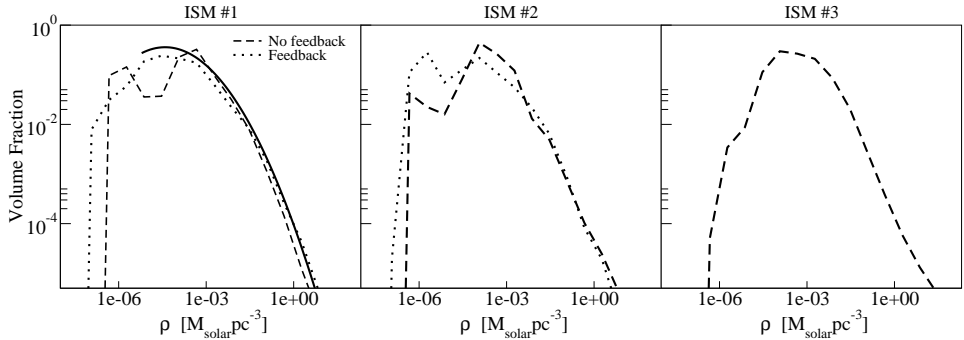}
\caption{PDF of the volume weighted gas density at 377\,Myrs for the three ISM types (running left to right). A log normal fit is shown as a solid line in the first panel.}
\label{fig:pdfs}
\end{figure}

The structure of these ISMs can be examined more quantitatively by looking at their 1D PDFs, as shown in Figure~\ref{fig:pdfs}. Comparing all three of the ISMs across the panels, we see a significant amount of substructure in the low density gas, but the high density end of the PDFs are surprisingly uniform. At these densities, all profiles are well fitted by a lognormal distribution. This remains true even when feedback is included.

\section{Observational comparison}

\begin{figure}
\includegraphics[width=6.7cm]{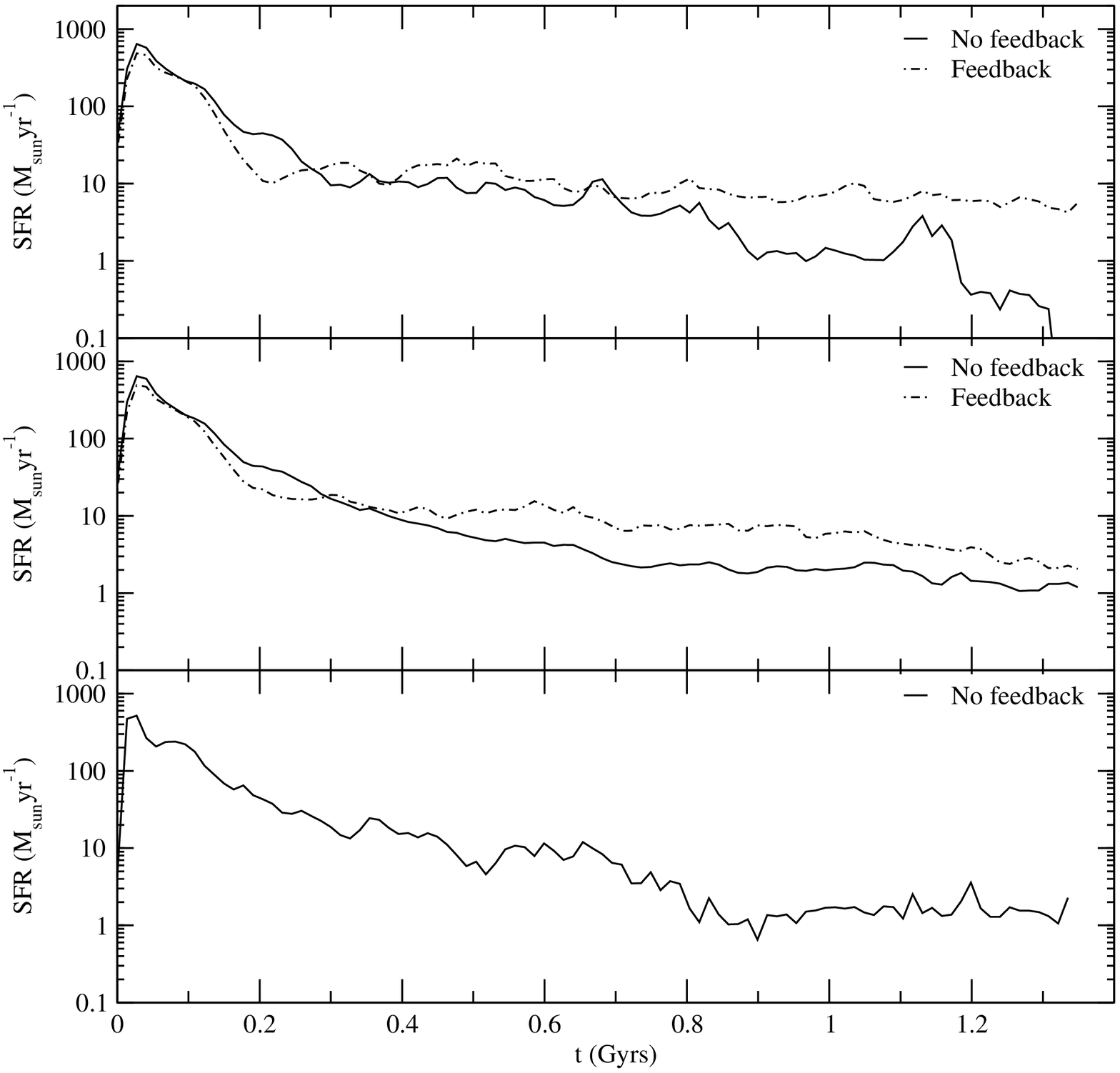}
\includegraphics[width=6.7cm]{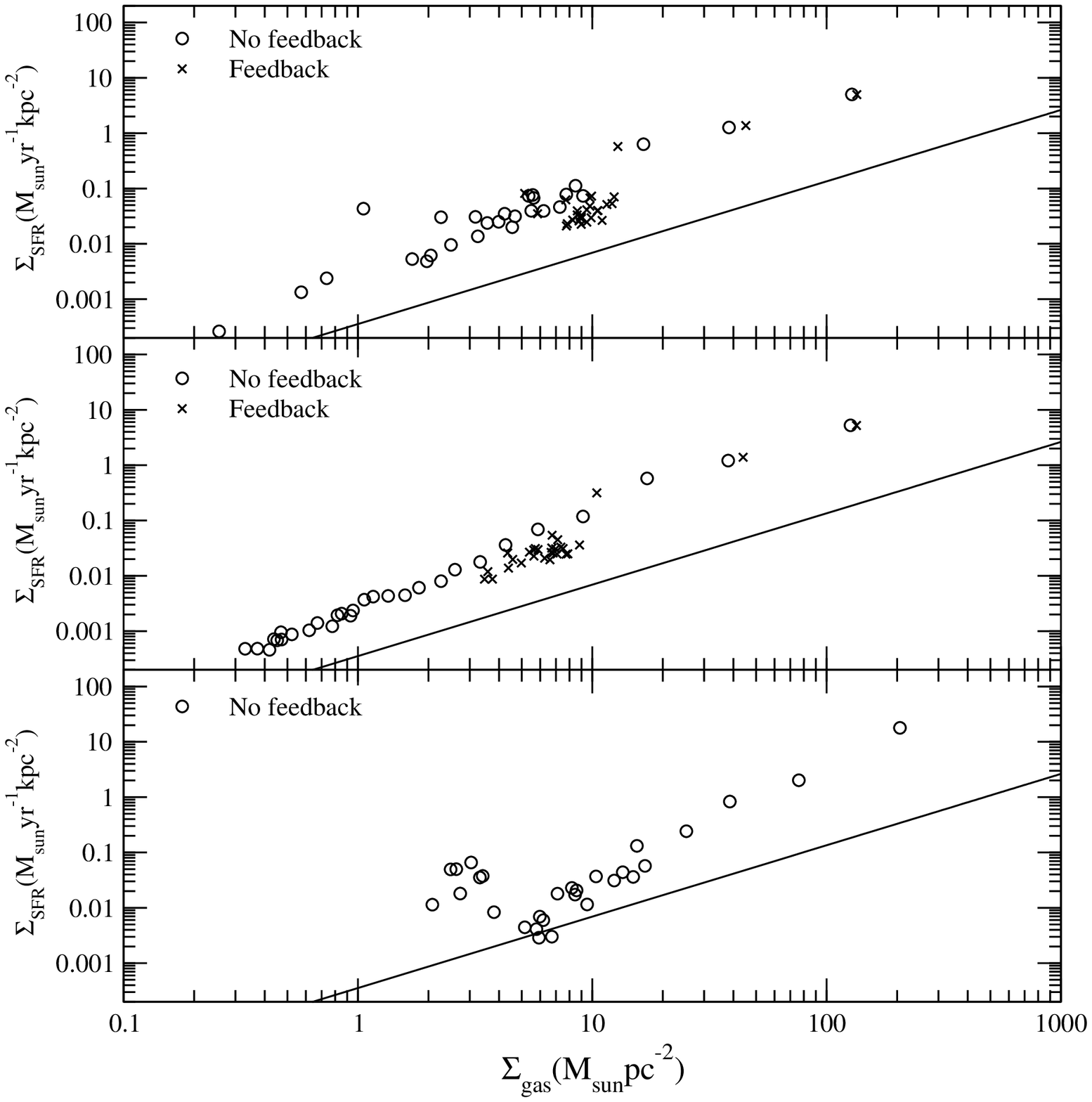}
\caption{The star formation history (left) and global Schmidt-Kennicutt law (right) for ISM models 1, 2 and 3 (top to bottom).}
\label{fig:obs}
\end{figure}

The left-hand plots in Figure~\ref{fig:obs} shows the star formation history for each galaxy disc. Without feedback, our ISM~\#1 disc converts all the available gas into stars, halting further star formation. By contrast, the addition of background heating quenches star formation by raising the temperature of the coldest gas and allowing the disc to show the beginning of self-regulation. The addition of feedback, however, is a much stronger effect, with the added energy destroying the star forming knots and thereby increasing the available gas at later times. The isothermal disc also shows a flattening in the star formation rate, but this is likely due to the confinement of the star formation to the denser parts of the disc which slows down the rate of gas consumption. 
We can also compare our results with the widely observed Kennicutt-Schmidt law \citep{Kennicutt1989}, as shown in the right-hand plot of Figure~\ref{fig:obs}. Here, we see that the observed gradient is reproduced well in both the first two ISM types, but less well in our isothermal disc. We do however, persistently overestimate the rate of star formation, even when feedback is included. This is likely due to our star formation recipe and is something to address in later work.

\section{Conclusions}

One of the most surprising results in this research is that despite significant structural differences in the ISM, the star formation properties remain largely unaffected. The reason for this can be seen in Figure~\ref{fig:pdfs}, where we see that the biggest differences between our three ISM types occur in the low to medium density gas. The high density, star-forming gas, meanwhile, forms a consistent lognormal profile in all cases. This result is largely positive; it suggests that a decent subgrid model for star formation can be used where a detailed ISM model is not possible, such as large-scale cosmological simulations. However, it also implies that a through understanding of star-formation cannot be found from the Kennicutt-Schmidt law, which is largely insensitive to the input physics. This work is presented in greater detail in \citet{Tasker2007}.

\end{document}